\documentclass[aps,pre,twocolumn,groupedaddress,amsmath,amssymb]{revtex4}
\usepackage{graphicx}
\usepackage{float}

\begin{document}

\title{Precursor-mediated crystallization process in suspensions of hard spheres}

\author{T. Schilling}

\affiliation{Theory of Soft Condensed Matter, University of Luxembourg, Luxembourg}

\author{H. J. Sch{\"o}pe, M. Oettel}
\affiliation{Institut f{\"u}r Physik, Universit{\"a}t Mainz, Germany}

\author{G. Opletal, I. Snook}
\affiliation{School of Applied Sciences, RMIT University, Melbourne Australia}


\begin{abstract}
We report on a large scale computer simulation study of crystal nucleation in hard spheres.
Through a combined analysis of real and reciprocal space data, a picture
of a two-step crystallization process is supported: 
First dense, amorphous clusters form which then act as precursors for the nucleation of well-ordered crystallites. This kind of crystallization process has been previously observed in systems that interact via potentials that have an attractive as well as a repulsive part, most prominently in protein solutions. In this context the effect has been attributed to the presence of metastable fluid-fluid demixing. Our simulations, however, show that a purely repulsive system (that has no metastable fluid-fluid coexistence) crystallizes via the same mechanism. 
\end{abstract}


\bibstyle{apsrev}

\maketitle

The crystallization process in complex fluids is not trivial.
For systems such as solutions of proteins, alkanes, and colloids it has 
been shown that crystal 
nucleation rates can be enhanced considerably if the supersaturated 
liquid is quenched to a state that lies close to a metastable fluid-fluid 
critical point \cite{Fibolo2005, Pan2005, Andersen2002, tenWolde1997, Talanquer1998, Shiryayev2004, Kashchiev2005, Lutsko2006}. 
The enhanced nucleation rate is generally attributed to the fact that 
the density fluctuations occuring in the vicinity of a metastable 
fluid-fluid critical point enable the system to evolve via a two-step process. 
First dense, amorphous precursors form and then the crystallization process 
takes place inside these. The prerequisite of this process scenario, 
the metastable fluid-fluid critical point, is easily realized in the 
systems listed above, which exhibit an interplay of repulsive and attractive 
interactions. 

However, it is worthwhile asking whether the two-step process occurs more 
generally. Surprisingly, there have been experiments indicating 
two-step crystallization occuring also in hard sphere systems, the simplest model system for 
liquids and crystals (see e.g.~Ref.~\cite{Schoepe2006}).
As the interaction energy between two 
hard spheres is either zero (no overlap) or infinite (overlap), the phase 
behaviour of the system is purely determined by entropy. In particular, 
for one component hard spheres there exists a stable crystalline phase but no
metastable fluid-fluid demixing region.

The crystallization kinetics in colloidal hard sphere systems has been 
studied experimentally using predominantly time resolved light 
scattering \cite{Schaetzel1992, Harland1995, He1996, Harland1997, Cheng2002, Schoepe2006, Schoepe2007} 
and to a lesser extend real-space imaging techniques \cite{Kegel2000, Gasser2001, Elliot2001, Prasad2007}. 
In the scattering experiments described in Refs.~\cite{Schoepe2006, Schoepe2007, Iacopini2009} the time-evolution of the structure factor has been interpreted 
using a two-step process model:   
In the induction stage precursors (compressed, structurally heterogeneous 
clusters) slowly grow. Then the precursors are converted into 
highly ordered crystals in a fast, activated process. 
In Ref.~\cite{Schoepe2006} it was suggested that size polydispersity 
limited growth is responsible for the induction stage. However, later it was 
argued that the precursor stage behaves 
in a similar fashion, regardless of polydispersity or of metastability 
suggesting that the precursor nucleation and the following conversion is not 
a special feature of polydisperse samples, but that it might constitute a 
fundamental process of crystal nucleation \cite{Schoepe2007}. 


The complete mechanism of precursor to crystal conversion is still unknown and 
difficult to obtain via structure factor analysis alone. 
A real space experiment would be highly desired. But as size 
polydispersity cannot be avoided in a lab experiment a high precision 
computer study appears to be the best choice. 

Hard spheres are easily realized on the computer. Simulations of hard sphere 
crystallization have been reported e.g.~in Refs.~\cite{Truskett1998, OMalley2003, OMalley2005} and of crystal nucleation kinetics e.g.~in Refs.~\cite{Gruhn2001, Auer2001}. One should bear in mind, however, that
nucleation is a typical example of a rare event, i.e.~an event 
that has a low reaction rate but high impact on the properties of a system.
Rare events, and in particular nucleation, are very often simulated by 
methods that are based on transition state theory \cite{Eyring1935}. The 
basic assumption underlying transition state theory is that the rare process 
can be reduced to the dynamics of ``slow'' variables which evolve in an 
effective free energy landscape formed by the ``fast'' variables (i.e. those 
that quickly adopt a Boltzmann distribution). 
The choice of slow variables already presupposes a certain dynamic scenario  
which does not necessarily hold in the experimental system. For instance,
the transition rates computed in \cite{Auer2001} using transition state theory
correspond to a one--step crystallization process and the results are
in disagreement with the experimental results.
Therefore we have carried out a computer simulation study which follows 
the nucleation kinetics directly (without using any biasing scheme that 
would require underlying assumptions on the nucleation pathway) and therefore 
allows for direct comparison to experiments.

We have performed Monte Carlo simulations of $N=216,000$ hard spheres 
in a box of volume $V = 59.2\times59.8\times59.2\,D^3$, 
where $D$ is the particle diameter. In the following we use the particle 
diameter as unit of length, $k_BT$ as unit of energy and attempted MC moves 
per particle (``sweeps'') as unit of time.
The system was prepared by a fast pressure quench from the stable liquid.
The subsequent crystallization dynamics were simulated at fixed 
$N$, $V$ and $T$ 
by small translational MC moves only -- a method which mimicks 
Brownian dynamics on long time-scales \cite{Chui2010}.
We used this approach, because it requires relatively little CPU time per 
move, a necessary property when running a simulation of such a large size. 
We let the system evolve for $10^6$ MC sweeps 
and sampled observables every 5,000 sweeps. 
The number density after the quench was $N/V = 1.03$ (volume fraction
$\phi=0.54$), 
a value which lies in the liquid-solid coexistence region close to the 
density of the solid at coexistence (ca.~$9\%$ above the coexistence 
density of the liquid). This corresponds to a chemical potential 
difference between the metastable liquid and the stable, almost completely 
crystalline state of $\Delta \mu \simeq -0.58$ per particle. 
Given that the interfacial 
tension is of the order of $0.5$ 
\cite{Davidchack2000,Davidchack.Morris.Laird:2006} 
one would expect the system to be far beyond the classical nucleation 
regime and to crystallize almost instantaneously. 
The self diffusion constant was 
$D_S = 2.3 \cdot 10^{-5} \frac{1}{\rm \#\, MC\, sweeps}$. 

We monitored crystallization by means of the q6q6-bond order parameter 
\cite{Steinhardt.Nelson.Ronchetti:1983,Wolde.RuizM.Frenkel:1995}.
For a particle $i$ with $n(i)$ 
neighbours, the local orientational structure is characterized by
\[
\bar{q}_{lm}(i) := \frac{1}{n(i)}\sum_{j=1}^{n(i)} Y_{lm}\left(\vec{r}_{ij}\right)\quad ,
\]
where $ Y_{lm}\left(\vec{r}_{ij}\right)$ are the spherical harmonics 
corresponding to the orientation of the vector $\vec{r}_{ij}$ between 
particle $i$ and its neighbour $j$ in a given coordinate frame.  
As we are interested in local fcc-, hcp- or rcp-structures, we consider $l=6$. 
We assign a vector $\vec{q}_{6}(i)$ to each particle, the elements 
$m=-6 \dots 6$  of which are defined as 
\begin{equation}
q_{6m}(i) := \frac{\bar{q}_{6m}(i)}{\left(\sum_{m=-6}^6|\bar{q}_{6m}(i)|\right)^{1/2}} \quad . \label{Defq6q6}
\end{equation}
Two neighbouring particles $i$ and $j$ were regarded as ``bonded'', if 
the dot product $\vec{q_6}(i)\cdot \vec{q_6}(j)$ exceeded $0.7$ (i.e.~if their 
local orientational order added up almost coherently). $n_b(i)$ is the number
of ``bonded'' neighbours of the $i$th particle.
Orientationally ordered clusters are defined as regions of bonded particles with $n_b(i)$ exceeding some common threshold value.   
In particular, clusters were called ``crystallites'' if $n_b > 10$ (i.e. almost perfectly hexagonally ordered).

\begin{figure}
  \centering
  \includegraphics[width=0.7\columnwidth]{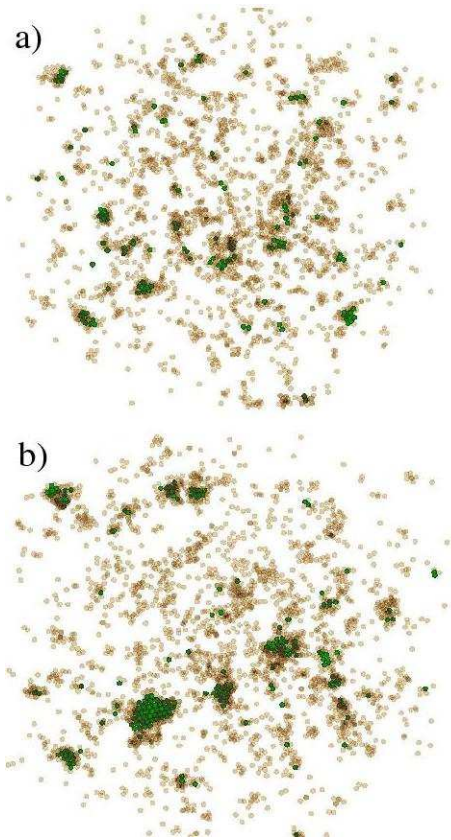}
  \caption{System snapshots at a) 350,000 sweeps and b) 450,000 sweeps. Particles with $n_b > 5$ (light brown) 
    and $n_b > 10$ (green). Particles with fewer bonds are not shown.}
  \label{fig:Snapshots}
\end{figure}

\begin{figure}
  \centering
  \includegraphics[width=1.1\columnwidth]{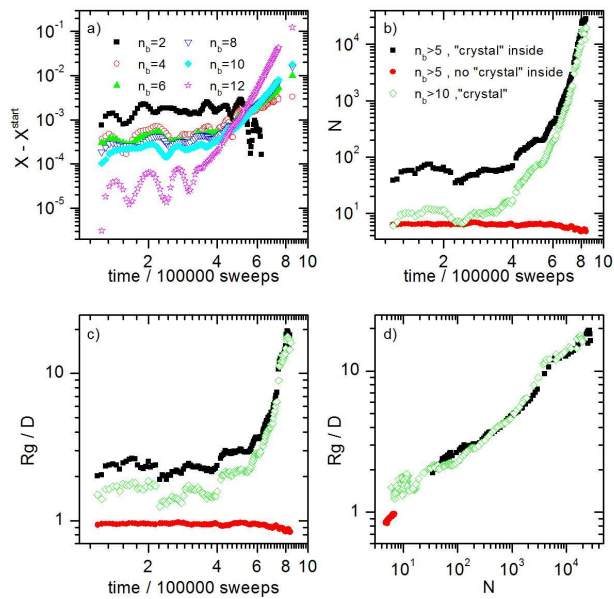}
  \caption{Evolution of crystallinity. a) Fraction $X$ of particles in clusters for various values of $n_b$ (number of crystalline bonds according to Eq.~\ref{Defq6q6} and text thereafter). b) Number of particles in crystallites (clusters with $n_b>10$, open green diamonds) and in low symmetry clusters ($n_b>5$) either containing crystallites (black squares) or not containing crystallites (red circles).  c) Radius of gyration of clusters (symbols as above). d) Radius of gyration versus number of particles in clusters.}
  \label{fig:AllRgNT}
\end{figure}

Fig.~\ref{fig:Snapshots} shows system snapshots at two times. Low symmetry clusters (LSC, $n_b > 5$) are light brown, crystallites ($n_b > 10$) are green. One can clearly see that the crystallites are nucleated inside the LSC. To quantify this effect Fig.~\ref{fig:AllRgNT} shows the evolution of crystallinity in more detail. In Fig.~\ref{fig:AllRgNT} a) the fractions of particles with given numbers of bonds $2 \le n_b \le 12$ are plotted. During an induction time of ca.~400,000 MC sweeps the amount of orientationally ordered material grows slowly. Then crystallization sets in, as shown by the evolution of particles located in pure fcc/hcp crystallites ($n_b=12$). The growth rates of all other ordered regions are markedly smaller, in particular for $n_b < 5$. 
Fig.~\ref{fig:AllRgNT} b), c) and d) show the evolution of crystallites ($n_b>10$) and LSC ($n_b>5$) where we distinguished between LSC that contain crystallites and LSC that do not. The average number of particles $N$ in (or likewise the radius of gyration $R_g$ of) the LSC is consistently larger than $N$ or $R_g$ of the crystallites, supporting the picture of crystallites growing inside LSC. Further analysis showed that the centers of mass of the crystallites and their surrounding LSC practically coincided.


\begin{figure}
  \centering
  \includegraphics[width=0.9\columnwidth]{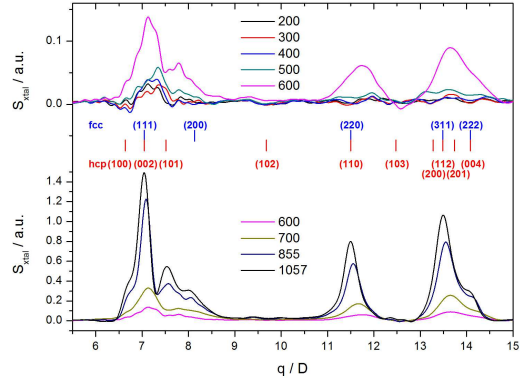}
  \caption{Evolution of crystal structure factor. Different colors correspond to
different times given in $10^3$ MC sweeps as indicated. For further details see text.}
  \label{fig:SqTime}
\end{figure}

To compare the simulation with light scattering data we extracted the radially averaged
pair correlation function as function of time. By Fourier transformation the structure factor of the system was calculated. Following the procedure first proposed by Harland and van Megen \cite{Harland1995} we obtained the time evolution of the crystalline structure factor (see Fig.~\ref{fig:SqTime}). 
The simulation results are very similar to the ones obtained in experimental 
investigations \cite{Iacopini2009}: At early times during the induction 
stage we observe only one broad peak close to the position of the fcc (111) 
peak stemming from 
the compressed precursor structures growing slowly in intensity while the 
width remains nearly constant - the scattering pattern changes quite slowly. 
After about 500,000 sweeps a structure factor stemming from compressed rhcp (random hexagonal closed packed) crystals can be identified, the higher 
order reflections are clearly 
visible: the crystallites evolve from the initial 
precursor structures to a rhcp structure. After the conversion is complete the 
structure does not change significantly during the main crystallization stage, 
but the intensity increases rapidly (600,000 sweeps till end). The peaks 
become smaller and shift to smaller $q$ values.

\begin{figure}
  \centering
  \includegraphics[width=1.0\columnwidth]{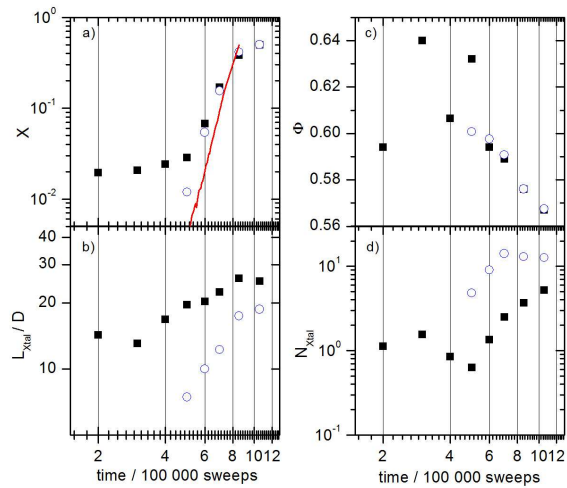}
  \caption{Evolution of parameters extracted from analysis of the crystal structure factor.
  a) Crystallinity, red line: data from Fig.~\ref{fig:AllRgNT}a ($n_b$=12) for comparison. b) average domain size. c) volume fraction and d) number of clusters/crystals. Black squares fcc: (111) peak, open circles: fcc (220) peak}
  \label{fig:X_etc}
\end{figure}

From the integrated area, the width and position of individual Bragg peaks 
the amount of crystallinity, the averaged domain size and the volume fraction 
of the clusters/crystals can be obtained. While the amount of crystalline 
material and the lattice constant can be determined with high accuracy, there 
is a noteworthy systematic error in the averaged domain size in our data 
analysis, pertaining to Fig.~\ref{fig:X_etc}.  In a polycrystal with rhcp structure the width of the peaks is 
connected with the crystal size distributions of crystals with different 
stacking parameters \cite{Kegel2000}. 
Especially the analysis of the hcp(002) (fcc(111)) 
reflection is difficult while the analysis of the hcp (110) (fcc(220)) peak 
is quite robust. As we have only a very small number of crystals in the sample 
the Scherrer formula to determine the crystal size is strictly speaking not 
fulfilled due to bad statistics. Nevertheless we show the determined 
quantities in Fig.~\ref{fig:X_etc} which can likely be compared in a 
qualitative way to experimental results \cite{Iacopini2009}.

In the induction stage the amount of orientationally ordered material and 
the size of the 
highly compressed precursors stays nearly constant ($2-4 \times 10^5$ sweeps). 
During conversion the precursors start growing while the increase in 
crystallinity is delayed and a significant drop in the number of precursors 
can be observed ($3-5 \times 10^5$ sweeps). During the main crystallization 
process also information stemming from the fcc(220) peak can be obtained 
($5-11 \times 10^5$ sweeps). Here crystallinity increases rapidly, crystal 
size and the number of crystallites show their strongest increase due to 
crystal growth and crystal nucleation. The crystals expand to reach the 
equilibrium volume fraction at the end of the crystallization process 
(which was not reached in these calculations). As discussed above the 
absolute values in the averaged crystal 
size stemming from the fcc(111) are prone to error leading 
to unphysically small values in the absolute number, however its time trace 
reflects the correct trend. 

We would like to emphasize that there is a remarkable resemblance of the time 
evolution of crystallinity extracted from the structure factor data with the
crystallinity fraction extracted from the maximally-bonded ($n_b=12$) crystallites
(see Fig.~\ref{fig:X_etc}a). This means that the orientation--averaged 
structure factor analysis can be mapped very well to the real-space analysis
of the mutual particle orientations. 

To summarize, we studied the nucleation process in hard spheres by combining a real-space 
bond-order analysis (typical for simulations and confocal microscopy 
experiments) with a reciprocal-space analysis of the time-evolution 
of the structure factor (typical for scattering experiments). 
Our simulations showed that nucleation in hard spheres is more 
complex than the traditional picture of one-step classical nucleation 
suggests. We identified the formation of 
dense clusters right after the quench containing particles 
that have high q6q6-coherence with at least half 
of their neighbours. The critical crystal nuclei observed in our simulation 
are not formed spontaneously in one step from random fluctuations but 
they appear inside these precursors of lower symmetry.  
The metastable fluid relaxes the density first, by producing dense low symmetry 
clusters, and later crystallites of perfect structure are formed. 

The two-step crystallization mechanism identified here for hard spheres is
akin to processes that have been observed in protein solutions and in 
suspensions of attractive colloids. Thus we conclude that metastable 
fluid-fluid demixing is not a necessary prerequisite. Different dynamics
for the two order parameters density and structure seem to suffice for
a two-step nucleation process. 

\begin{acknowledgments}
Financial support by SFB-TR6 and SPP1296 is gratefully acknowledged. 
TS thanks the NIC J\"ulich for CPU time. IKS would like to thank K.~Binder 
and T.~Palberg for their hospitality when visiting Mainz.
\end{acknowledgments}


\end{document}